\documentclass[intlimits,twoside,a4paper]{article}

\usepackage{xcolor}
\definecolor{ballblue}{rgb}{0.13, 0.67, 0.8}
\definecolor{OliveGreen}{rgb}{0,0.6,0}
\usepackage{graphicx}
\usepackage[cp1251]{inputenc}

\usepackage{xcolor}
\usepackage{soul}

\usepackage[eqsecnum]{cmpj3}

\issue{2017}{20}{3}{33802}
\doinumber{10.5488/CMP.20.33802}

\title{Contribution of cellular automata to the understanding of corrosion phenomena}
\author[M. Zenkri \textsl{et al.}]{M. Zenkri\refaddr{label1,label2,label3},
        D. di Caprio\refaddr{label2}, C. P\'erez-Brokate\refaddr{label4}, D. F\'eron\refaddr{label4,label5}, J. de Lamare\refaddr{label4}, A.~Chauss\'e\refaddr{label5}, F.~Ben Cheikh Larbi\refaddr{label1,label3}, F.~Raouafi\refaddr{label1,label3}}
\addresses{
\addr{label1} LPC2M, IPEST, Route Sidi Bou Said, B.P:51 2075 La Marsa, Tunisie
\addr{label2} Chimie ParisTech, PSL Research University, CNRS, Institut de Recherche de Chimie Paris (IRCP), Paris
\addr{label3} Facult\'e des Sciences de Bizerte, 7021 Jarzouna, Universit\'{e} de Carthage, Tunisie
\addr{label4} Den-Service de la Corrosion et du Comportement des Mat\'{e}riaux dans leur Environnement (SCCME), CEA, Universit\'{e} Paris-Saclay, Gif-sur-Yvette, France
\addr{label5} Laboratoire Analyse et Mod\'{e}lisation pour la Biologie et
l'Environnement, UMR 8587, CNRS-CEA-Universit\'{e} d'Evry Val d'Essonne, France
}

\date{Received April 29, 2017}

\begin{document}

\maketitle

\begin{abstract}
We present a stochastic CA modelling approach of corrosion based on
spatially separated electrochemical half-reactions, diffusion,
acido-basic neutralization in solution and passive properties of the oxide layers.
Starting from different initial conditions, a single framework
allows one to describe generalised corrosion, localised corrosion,
reactive and passive surfaces, including occluded corrosion phenomena as well.
Spontaneous spatial separation of anodic and
cathodic zones is associated with bare metal and passivated metal on the surface.
This separation is also related to local acidification of the solution.
This spontaneous change is associated with a much faster corrosion rate.
Material morphology is closely related to corrosion kinetics, which can be used
for technological applications.
\pacs 82.45.-h, 82.45.Bb, 07.05.Tp, 82.20.Pm
\vspace{-1mm}
\end{abstract}

\section{Introduction}
\vspace{-1mm}
This paper is dedicated to Jean-Pierre Badiali.

Corrosion represents a substantial loss of about 4\% of the world GDP yearly \cite{1,2}
and concerns most human activities.
Thus, understanding and modelling corrosion is of strategic importance for
technological applications ranging from engineering structures, transport
\cite{3} to high-tech biological implants \cite{4}.
Given its importance and the need for practical advances, two activities have
emerged in parallel with slightly different objectives.
``Corrosion engineering'' is aimed at practical advances and relies on collecting
data on corrosion in view of corrosion protection \cite{5} while
``Corrosion science'' being more academic, collects basic scientific knowledge and is aimed
at understanding the corrosion mechanisms \cite{6}.
In this paper, we contribute to the latter with some simulation modelling results.

Here, we shall study aqueous corrosion.
And we limit our purpose to simple electrochemical corrosion discarding other
processes like stress corrosion cracking, erosion, fatigue or microbial
corrosion \cite{7}.
Despite these restrictions, corrosion remains a complex phenomenon.
It involves chemical and electrochemical processes in  different phases like solid-liquid phases, and concerns phenomena in volume as well as interfaces.
A wide variety of processes are involved, such as electron transfer and species
diffusion \cite{8}.
All different phenomena involved have a cross-interaction and 
evolution details may depend on: the nature of the material, its forging history
or composition like in alloys.

Two main types of corrosion are observed: generalized corrosion
and localized corrosion. They have different features.
Generalized corrosion occurs when the deterioration of the metal leads to more or less
uniform thickness loss of metal on the surface, although at smaller length scale
there can be surface roughness~\cite{9}.
This type of corrosion is somewhat better understood or at least more predictable
since it can be characterised in terms of metal thickness loss per time unit.
On the contrary, localized corrosion is more tedious and may appear at apparently random locations
on the metal surface \cite{8}.
Although more limited in space and affecting a limited quantity of metal, its
consequences on the integrity of the device may be far worse.
In general, its kinetics is also much faster.
Studying localized corrosion has recently become more accessible due to the range of
new in-situ microscopic probes. For instance, electrochemical probes such as
scanning reference electrode technique (SRET) \cite{10} or scanning electrochemical
microscope (SECM) \cite{11,12,13}, physical probes like AFM, STM or chemical probes such as
focused ion beam-secondary ion mass spectroscopy (FIB-SIMS)\cite{14,15}.

In this paper, we will focus more specifically on localized corrosion
of metals in contact with a solution disregarding more complex
alloys.
In our opinion, three different aspects are  essential.
(\textit{i}) One important phenomenon related to localized corrosion is passivation.
One has to note that most metals are thermodynamically unstable in standard conditions such as aqueous environment,
a standard tool used to determine thermodynamic stability being the Pourbaix diagram \cite{16}.
Passivation phenomenon was introduced in 1790 by Keir \cite{17,18}.
After that, in the experiments of Faraday, it was shown how placing an iron sample
in a diluted nitric acid will increase its dissolution, but with increasing the concentration
of the acid, the sample is protected and its dissolution is stopped unless the surface is scratched
which removes the passive layer and the corrosion resumes \cite{19}.
Note that passivation is based on a paradox. The paradox states that the more
prone to corrosion a metal is, due to the formation of the passive layer, the
slower corrosion is.
(\textit{ii}) We have seen earlier that the local breakdown of passivation leads to localized corrosion
which corresponds to different corrosion rates at different points of the metallic surface.
This leads us to point out another specificity of corrosion, the fact that this is an electrochemical
reaction constituted of two half-reactions: the anodic and the cathodic reaction \cite{8}.
These two half-reactions can occur in the vicinity of one another but also at some distance.
This is typically the case of pitting corrosion where the anodic reaction is generally
inside the pit and the cathodic reaction is outside \cite{6}.
However, the two half-reactions are not independent. They require a charge transfer between the anodic and cathodic
half-reactions.
This transfer of electrons in the metal or oxide, and ions in the solution which must then be an electrolyte, is the so-called galvanic coupling.
(\textit{iii}) The existence of two distinct half-reactions carries another important feature
in the context of aqueous corrosion.
The  anodic and cathodic reactions are related respectively to
acidification or basification of the local environment \cite{20,21}.
This will be illustrated with the half-reactions presented in section~\ref{ssec:ChemReactAndCA}.

Experiments conducted in order to understand corrosion have been carried out for over two hundred years.
And they are still the focus of intensive studies, whether from the industrial viewpoint or also from
more academical understanding \cite{22}.
In parallel, theoretical approaches have also been developed.
They are clearly useful for understanding, explaining experimental results and  possibly lead to predictions.
In some cases, long term corrosion predictions modelling is the only option apart from
observation of archaeological analogues which do not
correspond to nowadays technology \cite{23}.

Different theoretical approaches have been developed.
One major challenge is that corrosion phenomenon starts at the atomic scale but
the resulting structure failures or influencing environmental conditions
are at the macroscopic scale.
The description of corrosion must then cover many different length scales as well as time
scales and a variety of different approaches can be used
\cite{24,25}.
They can be sorted from microscopic approaches (quantum
chemistry, atomistic DFT) to macroscopic approaches (fluid dynamics, mechanics)
or sorted by the subject like electrochemistry or thermodynamics.
Our approach can be understood in a scale hierarchy.
At the macroscopic scale, there are, for instance, approaches based on differential
equations such as finite element methods (FEM) or its variants FDM (finite
difference methods), BEM (boundary element methods)
\cite{24,26,27,28} but these
are essentially deterministic \cite{29}.
At the other end of the scale there are approaches dealing with the atomistic or molecular
aspects \cite{30,31,32} which are often
interested in the initiation of corrosion and particularly in the breakdown of the passive layer.
Other approaches are at an intermediate scale. These are the mesoscopic
approaches, some still related to the microscopic scale like KMC (kinetic Monte Carlo) approaches
\cite{33} which are still atomistic but on larger scales while some
are distinctly at a larger scale like cellular automata (CA) approaches (see for example reference \cite{34} for an overview).

As pointed out, no modelling can be comprehensive regarding the corrosion phenomenon.
Such modelling should span across so many spatial scales and embrace a large number of phenomena.
In our case of interest, passivation breakdown occurs at the atomic level whereas diffusion, for instance,
takes place at mesoscopic to macroscopic scale.
We choose here to focus on a mesoscopic scale approach which, in our opinion, provides an interesting
link between microscopic and macroscopic, more precisely, CA approaches.
These approaches have been used to describe various aspects of corrosion \cite{34}.
One main contribution of this type of approaches is to couple
chemical and electrochemical aspects to the evolutions of morphological
aspects \cite{35,36,37}.
First applied to generalized corrosion, they have been used for
studying the pitting corrosion \cite{38}.
The microscopic features cannot be modelled in detail at the mesoscopic level. However, they can be
introduced via some stochastic evolution rules which will depend on the probabilities which
represent the atomistic processes that can induce a passivity breakdown
at a given location.
This stochastic element is essential and contrasts with the macroscopic deterministic approaches.

The model we consider accounts for passivity, separates anodic and cathodic
reactions and couples the half-reactions on the metal side and on the
solution side. We will emphasize the relation between the kinetics and morphological aspects
properties.
These are essential features of localized corrosion.
The model was first developed by Jean-Pierre Badiali and collaborators
within a two dimensional description \cite{39,40}.
It has recently been generalized to three dimensions
\cite{41,42}.
In this paper, in section~\ref{sec:CAmodel}, we present the model.
The model is then applied to three different case studies which have been previously
investigated. We will emphasize common features and specificities, in section~\ref{ssec:CaseStudies}.
Finally, in section~\ref{ssec:newmaterials}, we will present new applications and discuss some perspectives.

\section {Cellular automata model of corrosion}\label{sec:CAmodel}

\subsection{Why and which cellular automata modelling}
Before presenting the model, we would like to stress our tribute to
Jean-Pierre Badiali's rationalisation of corrosion.
It was in his mind that corrosion is a highly complex process
and tackling all aspects is clearly out of reach at present.
In this context, he thought it would be beneficial to forge a simple model, which
he chose within the CA framework, to help analysing and understanding.

The CA allows us a simple lattice representation of reality, associating lattice sites to the states of matter.
These states evolve following simple rules. These rules can represent both
changes in the states of matter, i.e., chemical reactions, but also diffusion (swapping algorithm).
It has recently been shown that in contrast to common belief, CA can provide
a quantitative description of electrochemical processes \cite{44}.
CA is appealing taking into consideration that combining spatial representation and evolution rules, it
naturally correlates reactivity to local environment, hence morphology, and
provides mutual feedback mechanisms which can drastically modify the kinetics.
The approach uses an intermediate mesoscopic scale which can provide large-scale effects
of a microscopically randomly generated corrosion event and can cover far longer corrosion
times than those that can be simulated at atomistic or molecular scale.

A first illustration of this mutual effect is seen in early works of Jean-Pierre
Badiali on lithium batteries, illustrating the fundamental role of
passivation on the morphology and kinetics
\cite{45,46,47,48}.
The initial point was the Eden model that used to study the morphology of the growth of the passive layer on the lithium
immersed in a solvent. They observed a porous morphology of the passive film
whose features depended on chemical, geometrical, deposition rate parameters. In this modelling, they introduced the
idea of lattice sites poisoning \cite{45}.
This started a series of models on passivation \cite{46,47,48}.

The second class of phenomena that affect the morphology and kinetics is that of diffusion.
A few years later in 2001, Taleb et al. studied the formation of films
separating two interfaces: metal|layer and layer|ionic solution \cite{50}.
This model considers: (\textit{i}) metal corrosion at metal|passive layer interface,
(\textit{ii})~layer growth at the passive layer|solution interface
and (\textit{iii}) diffusion of metallic species from the metal towards the passive layer|solution
interface to transform into layer and contribute to its expansion.
The novelty lies in the coupling of the two interfaces due to the diffusion species.
This study was compared to the results obtained from this extension to the Eden model.
In the Eden model, the species instantly reach the interfaces at any position
which would correspond to infinite diffusion speed and there is not size limit
to the growth. The new model entails a hindrance from diffusion and
shows how the morphology of interfaces and kinetics is tributary from diffusion.
This model first studied in \cite{50} was modified changing the Pilling-Bedworth
parameter (ratio of the oxide to metal molar volume) to a typical value
larger than one \cite{51,52}.
In \cite{53}, it was shown that such a model with the addition of different
kinetics according to localization within the pit could predict a large variety of morphologies.
The initial focus of the models was on the solid layer growth with variants
accounting for the effect of the defects in the metal \cite{54} and
of dissolution of the passive layer on the evolution processes \cite{55,56}.

Finally, these initial models assumed an anodic and a cathodic reaction at the same location.
Then, an important step accomplished by Jean-Pierre Badiali and coworkers was to
consider the two electrochemical half-reactions at different
locations \cite{39,40}.
Doing so, the description of the solution also had to be changed since acidic species are
created at the anodic reaction and, vice versa, basic species are created at the cathodic site.
This new model naturally introduced heterogeneity in the description, on the solid surface
as well as in solution.
The heterogeneity leads to different kinetics at different locations of the metal surface
and induces roughness of the corrosion front and possible detachment of small
islands with an undefined shape from the passive layer; the chunk
effect. Later in 2007, they related the deviation of the Faraday's law with the chunk effect
\cite{39,40,56}.

All these works have been essential steps in improving our understanding of
corrosion. In particular, the more recent model with separate anodic and cathodic
reactions provides effective scenarios of corrosion.
To allow for comparison with experimental observations, the model has recently
been generalized to three dimensional space
\cite{41,42}.
We now give the details of this model.

\subsection{Chemical reactions and CA model} \label{ssec:ChemReactAndCA}

The physico-chemical model has been presented before in
\cite{41,42}
and is summarized in table~\ref{table:electrochemicalreactions}.
For aqueous corrosion, regardless of the nature of the metal, the model is based on
simplifying assumptions: basic electrochemical reactions (anodic and cathodic)
and simplified chemistry (no pollutants, only H$^+$ and OH$^-$ ions).
The reactions at the metallic surface are simultaneous anodic and
cathodic half-reactions that may occur at two randomly distributed sites of the
surface, they have been detailed in \cite{41}.
In the case of aqueous corrosion, these are redox electrochemical reactions,
possibly followed by hydrolysis.
We consider the case of hydrogen evolution.
They are summarized in the first column of table~\ref{table:electrochemicalreactions},
with two alternative realizations depending on acidic or basic environment.
The two half-reactions are called spatially separated (SSE) reactions.
As mentioned above, they are not independent.
They require a flow of electrons between the anodic and the cathodic sites as well as
ions in the solution.
The model must then ensure a continuous metallic path to allow for electrons to travel
between anodic and cathodic sites.
We also account for local electrochemical reactions in the form of a passivation reaction
in the basic environment and a dissolution reaction in neutral or acidic environment
named Spatially Joint reactions~(SJ).
In addition to the surface reactions, we also define  the neutralisation reaction in solution
      \begin{align}
      \mathrm{H^+ + OH^- \longrightarrow H_2O}.
      \label{eq:neutralization}
      \end{align}

\begin{table}[!b]
\caption{(Color online) Electrochemical reactions summary.}
\vspace{2ex}
        \centering
\footnotesize
        \begin{tabular}{|c|c|l|l|}
        \hline\hline Chemical reaction & CA rule &
        Environment & Probability \\
        \hline\hline \multicolumn{4}{|l|}{Anodic} \\
        \hline $\mathrm{\textcolor{red}{M} + H_2O \longrightarrow MOH_{aq} +
\textcolor{OliveGreen}{H^+} + e^-}$ &
\textcolor{red}{Reactive}
        $\longrightarrow$ \textcolor{OliveGreen}{Acidic} & Acidic,
Neutral & $P_{\mathrm{sse}}$ \\
         $\mathrm{\textcolor{red}{M} + \textcolor{blue}{OH^-} \longrightarrow
\textcolor{violet}{MOH_{solid}} + e^-}$ &
\textcolor{red}{Reactive} $+$
         \textcolor{blue}{Basic} $\longrightarrow$
\textcolor{violet}{Passive} $+$ Neutral & Basic & $P_{\mathrm{sse}}$ \\
        \hline \multicolumn{4}{|l|}{Cathodic} \\
        \hline $\mathrm{\textcolor{OliveGreen}{H^+} + e^- \longrightarrow \frac{1}{2} H_2} $ & Surface +
        \textcolor{OliveGreen}{Acidic}$\longrightarrow$ Surface + Neutral & Acidic, Neutral & $P_{\mathrm{sse}}$ \\
         $\mathrm{H_2O +e^- \longrightarrow \frac{1}{2} H_2 +
\textcolor{blue}{OH^-}} $ & Surface + Neutral $\longrightarrow$ Surface + \textcolor{blue}{Basic}& Basic & $P_{\mathrm{sse}}$\\
        \hline \multicolumn{4}{|l|}{Spatially Joint (SJ) reactions} \\
        \hline
         $\mathrm{\textcolor{red}{M} + H_2O \longrightarrow
\textcolor{violet}{MOH_{solid}} + \frac{1}{2} H_2 } $ &
\textcolor{red}{Reactive}$\longrightarrow$ \textcolor{violet}{Passive}& Basic & 1 \\
         $\mathrm{\textcolor{violet}{MOH_{solid}} \longrightarrow MOH_{aq} } $ &
\textcolor{violet}{Passive}$\longrightarrow$ Neutral & Acidic, Neutral & $P'_{\mathrm{oxi}} \times
N_{\mathrm{exc}}$,
 $P_{\mathrm{oxi}}$ \\
 \hline\hline    \multicolumn{4}{l}{Surface = Reactive or Passive,
   $N_{{\mathrm{exc}}}$ = Algebraic
 difference of acidic and basic sites around a selected metal site} \\
        \end{tabular}
\label{table:electrochemicalreactions}
\end{table}

\begin{figure}[!b]
\centerline{\includegraphics[width=0.5\textwidth]{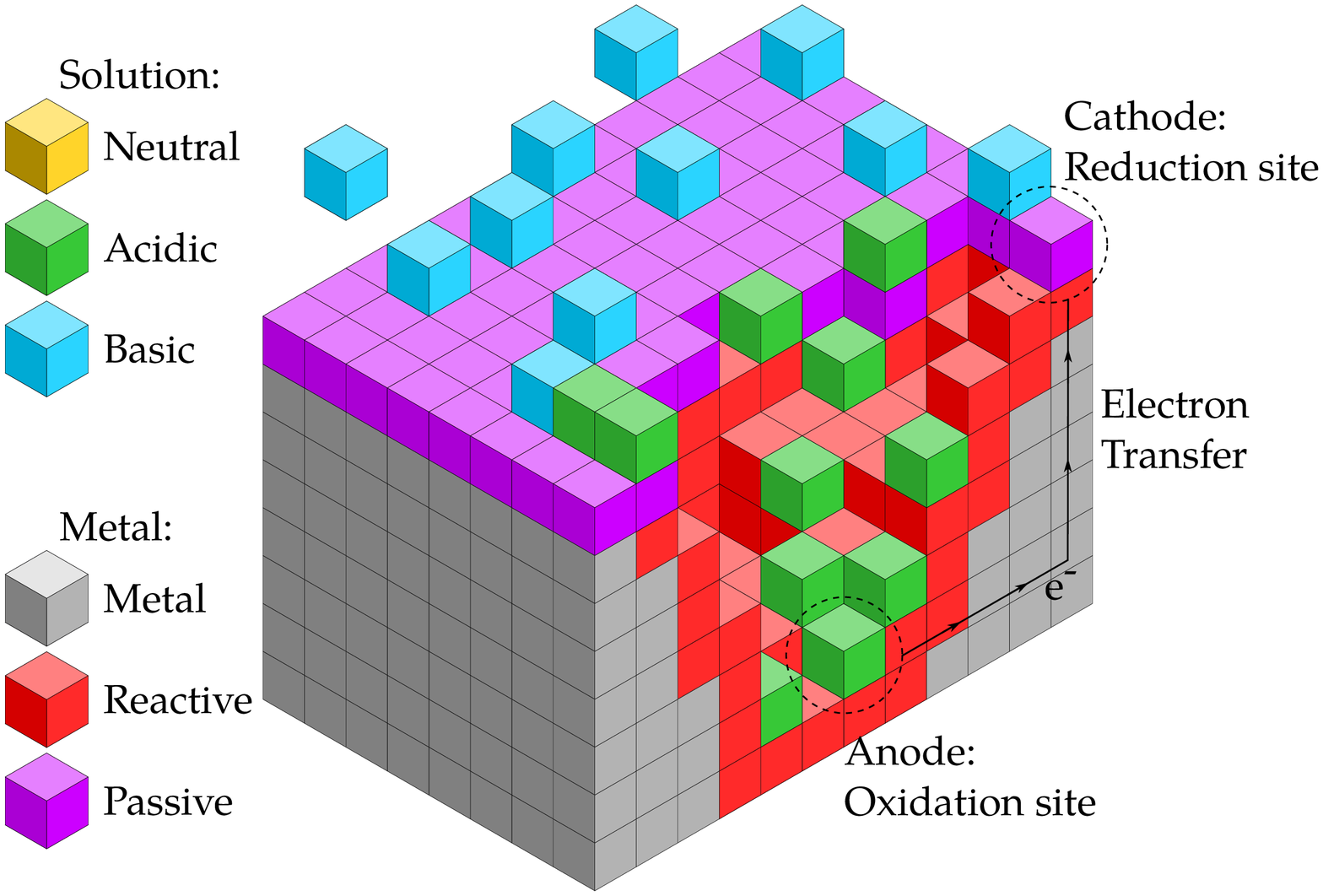}\hspace{8ex}\includegraphics[width=0.25\textwidth]{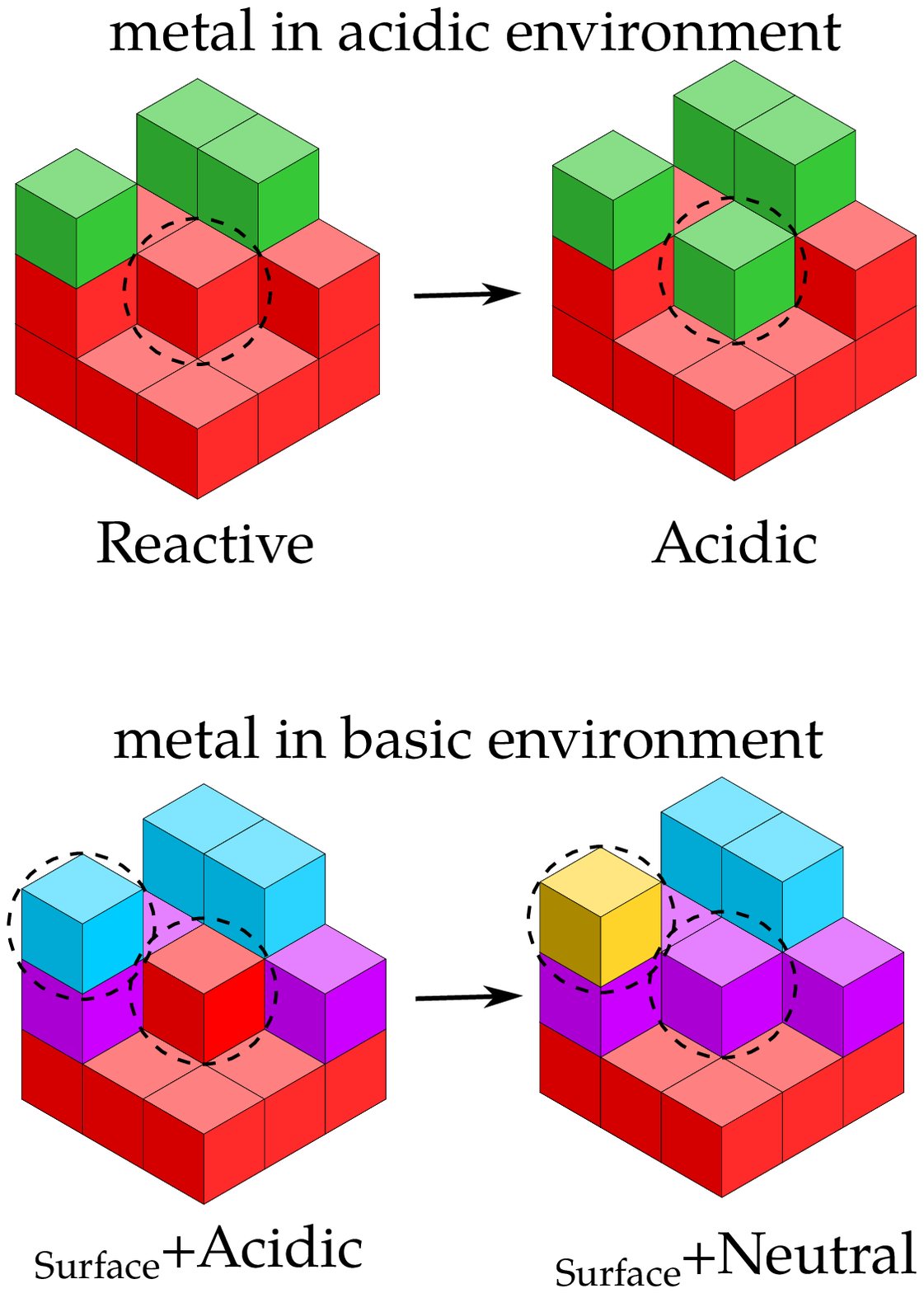}}
\caption{(Color online) Schematic representation of the system in the CA representation on the left and
typical surface reactions according to the local acidity/basicity environment on
the right.}
\label{fig:modelca}
\end{figure}

The counterpart of the physico-chemical system in the framework of CA has also been presented in
\cite{41,42}
and shown in the second column of table~\ref{table:electrochemicalreactions}.
The lattice is cubic and each lattice cell is given a state, the typical lattice size used here
will be $256\times 256\times 256$ sites.
For the solid species, we define site M for the bulk sites of metal, R for metal
reactive sites on the surface and  P for the passive layer constituted of oxide.
The species in solution are A for acidic $\mathrm{H}^+$ ions and B for basic $\mathrm{OH}^-$ ions
and finally the neutral solution is E.
The CA is given the Moore connectivity with 26 neighbours, to allow for a finer range of acidity
values and less restrictive diffusion displacements.
The reliability of the reaction-diffusion algorithm has been verified in \cite{44}.
As indicated in table~\ref{table:electrochemicalreactions}, some reactions are associated with the
probability parameters $P_{\mathrm{sse}}$ for SSE reactions, $P'_{\mathrm{oxi}}$ and $P_{\mathrm{oxi}}$
for dissolution reaction in acidic and neutral media, respectively, and with the
variable $N_{\mathrm{exc}}$
accounting for the local pH.
A typical configuration and surface reactions are schematically shown in figure~\ref{fig:modelca}.

The neutralization reaction [equation~(\ref{eq:neutralization})], in the CA, reads
      \begin{align}
      \mathrm{A + B \longrightarrow E}\,,
      \label{eq:neutralizationCA}
      \end{align}
and the diffusion is represented in the CA by a swapping procedure, for instance the acid A
moves from position $1$ to position $2$ in a random direction and neutral solution swaps
in the opposite direction:
      \begin{align}
      \mathrm{A_1 +E_2 \longrightarrow E_1 + A_2}\,,
      \label{eq:diffusion}
      \end{align}
where $\mathrm{A}$ can be replaced by $\mathrm{E}$ and the position indices interchanged.
This corresponds to a random walk which follows the relation $\langle l^2 \rangle = 2dD t$,
where $\langle l^2 \rangle$ is the average square distance covered in a time $t$,
$d=3$ is the space dimension and $D$ is the diffusion coefficient of the species, which
for simplicity is taken identical corresponding to the average value for $\mathrm{H^+}$, $\mathrm{OH^-}$
and equal to $D = 7.3\times10^{-5}$~cm$^2$s$^{-1}$ \cite{57}.
The typical dimension of the lattice site will be $a=10$~{\textmu}m, for which
we obtain the simulation time step $\Delta t_{\mathrm{diff}} = 9.48\times10^{-3}$~s.
This value takes into account a factor 2 due to the fact that in the algorithm, the swapping procedure is taken
for both configurations of position indices.

A more detailed description of the algorithm has been given as in \cite{41,42}.
There is a main corrosion loop with an inner diffusion, which is executed
$N_{\mathrm{diff}}$ times.
$N_{\mathrm{diff}}$ parameter then sets the corrosion time step
according to $\Delta t_{\mathrm{corr}} = N_{\mathrm{diff}} \Delta t_{\mathrm{diff}}$.

The electric connectivity condition between anodic and cathodic sites on the main metal
piece is ensured using the burning algorithm which has been adapted for a
parallel algorithm on GPU computing \cite{41,58}.
The roughening of the corrosion front can induce metal detachments. Now disconnected
from the main piece of metal, such islands no longer react in the algorithm.
To have these islands evolving, a dissolution mechanism is introduced with
probability $P_{\mathrm{diss}}$. In this paper, $P_{\mathrm{diss}}=1$. This implementation of the algorithm
does not essentially modify the dynamics of the main corrosion front.

\section{Results} \label{sec:Results}

\subsection{Case studies} \label{ssec:CaseStudies}
\begin{figure}[!h]
\centerline{\includegraphics[width=0.99\textwidth]{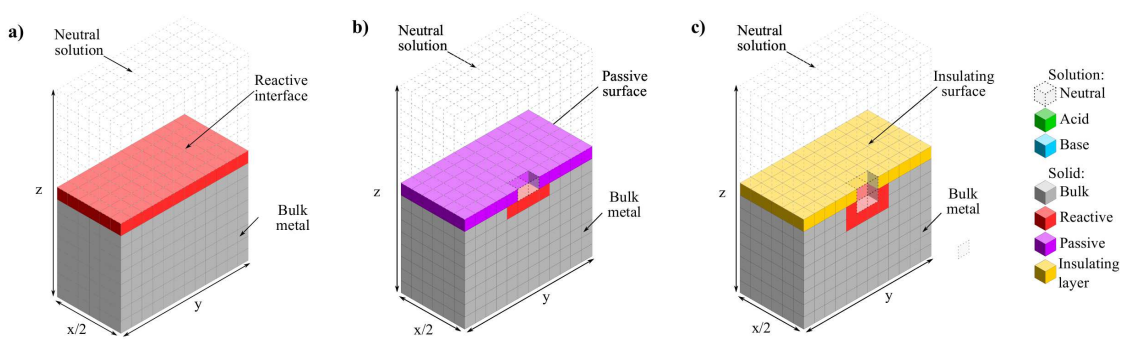}}
\caption{(Color online) Section of the starting configuration with  a) bare metal surface, b) passivated surface with a default and
c) insulating surface with a default.}
\label{fig:initialCA}
\end{figure}

In what follows, we recall three case studies which have been recently investigated
pointing out common features and differences.
For comparison, these case studies are investigated with
identical parameters, they will only differ by their initial state.
We thus consider three initial configurations corresponding to the bare
metal, a metal passivated with a small default on the passivating layer, e.g., a
passivated metal with a scratch and finally a metal covered with an insulating
layer, e.g.,  a metal with a layer of paint on top again with a small default.
The systems are shown in figure~\ref{fig:initialCA}.
The parameters have been chosen to simplify the system and retain only the main phenomena.
The parameters are $N_{\mathrm{diff}}=200$,  $P_{\mathrm{sse}}=0.5$, $P_{\mathrm{oxi}}=0$,  $P'_{\mathrm{oxi}}=5\cdot 10^{-4}$.
Under these conditions, not all SSE reactions are considered but the
depassivation due to dissolution of the passive layer is taken into account with
the non-zero value of $P'_{\mathrm{oxi}}$.
For computing time reasons, the value of $N_{\mathrm{diff}}$ is relatively small and corresponds
to rather high corrosion rates or anodization conditions, but it is shown in  \cite{42}
how to extrapolate to more realistic values.
We now briefly describe a common scenario, of which realization will be specific
for the case studies.

\subsubsection{Bare metal surface}

\begin{figure}[!t]
\centerline{\includegraphics[width=0.99\textwidth]{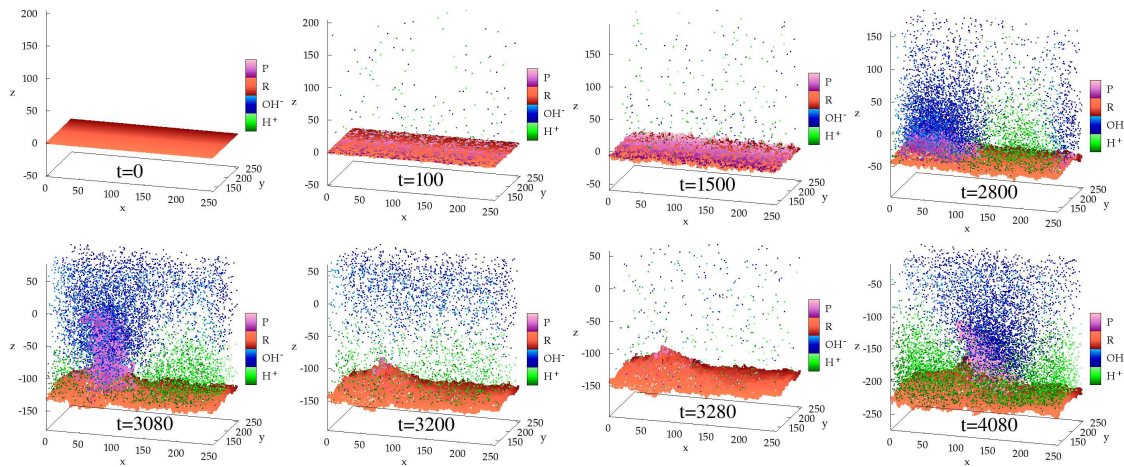}}
\caption{(Color online) Evolution of the corrosion front from an initially bare metal for time steps
indicated in the figure.
Note that periodic boundary conditions are assumed in the horizontal plane.}
\label{fig:generalisedsnap}
\end{figure}
Figure~\ref{fig:generalisedsnap} presents snapshots of the evolution of the system.
The initial system corresponds to a bare metal surface in contact with the
solution, $t=0$.
The corrosion is initially slow and the system follows a generalized corrosion regime $t=100, 1500$.
The flat surface is gradually covered with oxide and remains relatively uniform
although with some small roughness and the solution is homogeneous with initially few
acidic and basic species.
Acidic and basic species are produced respectively at the anodic and cathodic
sites and then they diffuse in a solution and neutralise each other when they meet one another.

\begin{figure}[!t]
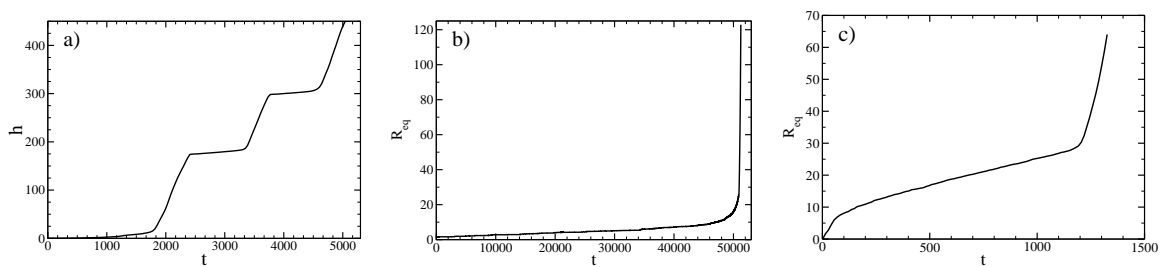

\centerline{\includegraphics[width=0.31\textwidth]{figure06a}\hspace{2.4ex}\includegraphics[width=0.317\textwidth]{figure06b}\hspace{2.4ex}\includegraphics[width=0.33\textwidth]{figure06c}}
\caption{Material loss as a function of time shown as equivalent a) height $h$
for generalized corrosion, or for pitting b)
and occluded c) cavity radius $R_{\text{eq}}$ as a function of the number of time steps $t$.}
\label{fig:materloss}
\end{figure}

From this uniform situation, a gradual increase of corrosion can be seen on the
average height~$h$ figure~\ref{fig:materloss}~(a), increasing $h$ is oriented in the direction
of corrosion.
Faster corrosion is correlated with an increase in concentration of acidic and basic species.
The stochastic simulation may give rise to inhomogeneities of acidic and basic
species which for a given diffusion parameter $N_{\mathrm{diff}}$ may not
have the possibility to neutralize.
The mechanisms in the model show that this inhomogeneity is correlated with the
surface state via the surface half-reactions.
More precisely, following the thermodynamics, the model assumes that the passive layer (oxide layer)
is stable in the basic environment. This means that such a region will host cathodic
reactions leading to a further basification of the environment.
The neighbouring cathodic region will, in contrast, remain acidic sustaining the corrosion
evolution since the oxide layer is not stable in acidic conditions.
This results in \textit{autocatalytic} evolution of the instability which maintains
the excess of one species in a given region, and the corrosion rate increases.

In the case of the plane metal surface investigated, the emergence of a fully passivated
region associated with a basic environment is associated around that zone to an acidic
zone where fast corrosion takes place ($t=2800$). As the corrosion around proceeds, this
leads to the formation of a peninsula ($t=3080$). As it grows, this peninsula is corroded around
its base and may finally detach ($t=3200$). It is dissolved at once due to the
value of $P_{\mathrm{diss}}=1$.
This metal detachment is the chunk effect: this phenomenon was originally observed by Thiel and Eckel \cite{59} and subsequently in
\cite{60,61,62},
especially on magnesium. More recently, it has shown a renewed interest in biodegradable
materials \cite{63,64}.
This metal detachment sets the surface back to a state close to the initial state ($t=3280$)
and the same scenario takes place in cycle with a new formation of a peninsula ($t=4080$).
This cyclic behaviour is clearly seen in the material loss, figure~\ref{fig:materloss}~(a).

\subsubsection{Pitting}

In the case of pitting, the initial surface is passive and corrosion should 
 not take place, with the exception 
 of the small default in the centre of the simulated surface. An other default 
 may possibly materialize 
 by dissolution of the passive layer but the event is rare as the probability $P'_{\mathrm{oxi}}=5\cdot10^{-4}$ is very small.
\begin{figure}[!t]
\centerline{\includegraphics[width=0.99\textwidth]{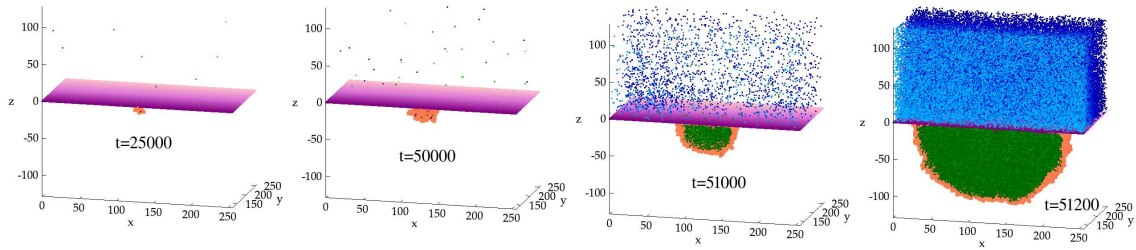}}
\caption{(Color online) Evolution of the pit front from an initially passivated metal with a default
in the cover at times indicated in the figure. Colour code as in figure~\ref{fig:generalisedsnap}.}
\label{fig:pittingsnap}
\end{figure}

At the passive layer default, a cavity grows slowly.
In the cavity, one may have anodic or cathodic reactions, whereas on the outer
surface, only cathodic reactions can take place.
This initial situation is asymmetric and favours the corrosion inside the pit
due to the creation of acidic sites in the cavity.
However, the growth is slow to start with, since the pit is small and the number of
acidic sites created by the anodic reaction is small.
The cavity does not evolve much from the initial system through $25000$ and
$50000$ time steps, as can be seen in figure~\ref{fig:pittingsnap}.
Note that the degradation of the passive layer is slow, it is protected
by the basic environment on the outside, and the species inside
the cavity remain trapped.
As the cavity gradually grows, the number of acidic sites increases
and may reach a sufficient accumulation in the pit for
the autocatalytic scenario to start.
Corrosion dramatically increases in this phase visible in the snapshot at
$t=51000$ and the cavity covers the simulation lattice only $200$ time steps afterwards.
This behaviour is clearly observed in figure~\ref{fig:materloss}~(b), showing the
equivalent cavity radius calculated as the number of material loss sites
$N_{\mathrm{loss}}$ according to  $R_{\text{eq}} = (\frac{3N_{\mathrm{loss}}}{2\piup})^{1/3}$.
Three orders of magnitude separate the two corrosion regimes. %2040

\subsubsection{Occluded corrosion}

\looseness=-1 This case is the three dimensional extension of a case previously studied
in two dimensions by Jean-Pierre Badiali and coworkers
\cite{39,45,46}
and has been investigated in \cite{41}.
The starting point for the system is represented in the schematic CA
representation given in figure~\ref{fig:initialCA}~(c).  A new species is added here
which represents the insulating layer (e.g., a paint) under which the occluded
corrosion takes place.
\begin{figure}[!b]
\centerline{\includegraphics[width=0.9\textwidth]{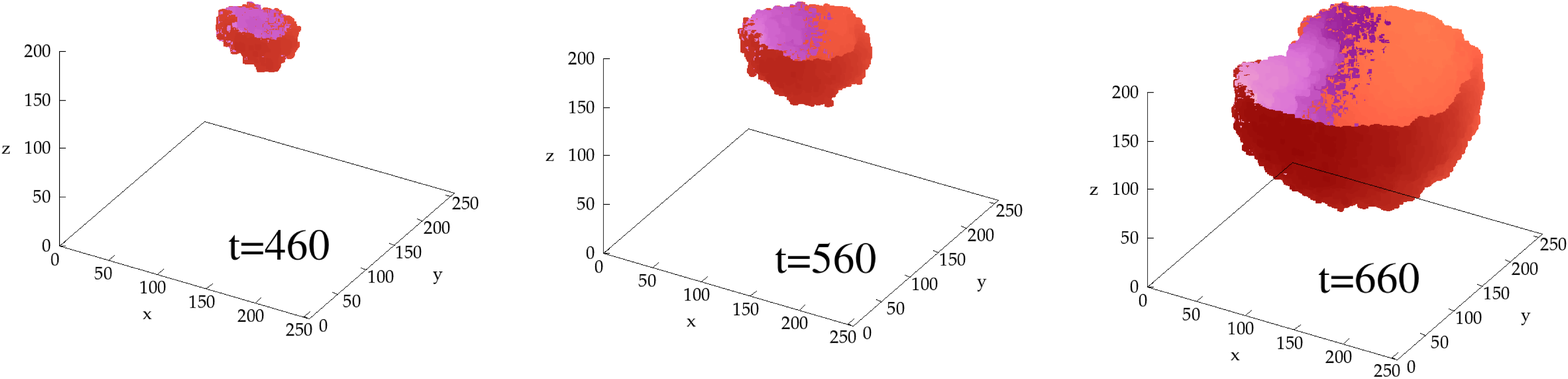}}\vspace{2ex}
\centerline{\includegraphics[width=0.9\textwidth]{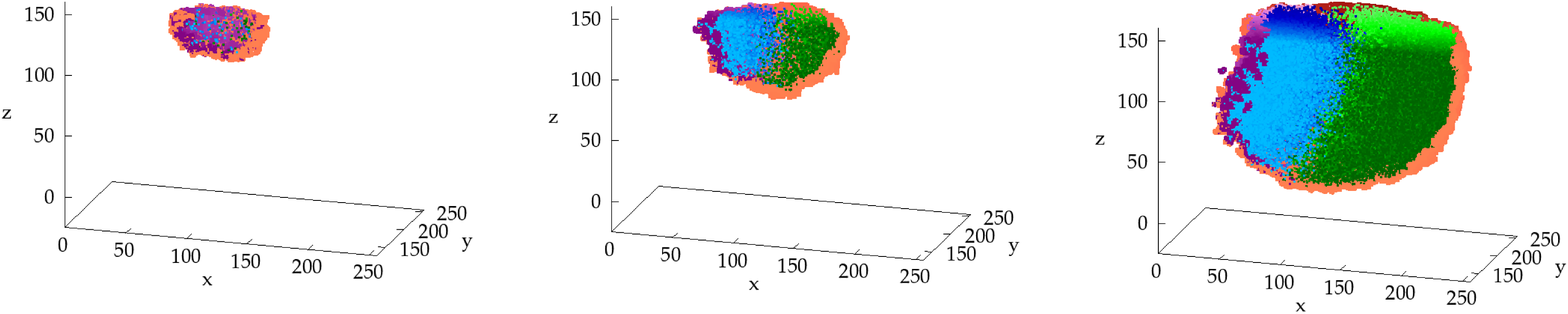}}
\caption{(Color online) Occluded corrosion snapshots, from left to right for simulation times as indicated.
The insulating layer and the bulk metal are not shown.
Top perspective with only reactive metal and passive
layer shown. Bottom side view also showing acidic and basic species.
Colour code as in figure~\ref{fig:generalisedsnap}.}
\label{fig:occludedsnap1}
\end{figure}
In figure~\ref{fig:occludedsnap1}, one can observe snapshots of the system.
Corrosion is slow to start with, at $t=460$, the cavity is small, the passive layer
appears distributed over the whole cavity surface as can be seen in the upper
and lower representation. There are also few acidic and basic species in the cavity.
Figure~\ref{fig:materloss}~(c) shows the evolution of the average radius of the cavity.
The slow corrosion regime is a generalized corrosion regime.

This regime is followed by a localized corrosion regime. This is visible in the fast evolution
of the cavity radius, figure~\ref{fig:materloss}~(c).
The cavity is asymmetric.
There are now two distinct regions in the cavity visible in figure~\ref{fig:occludedsnap1} for $t=560, 660$. One region is passivated, rather rough
and experiences little material loss. It is surrounded by a basic solution.
The other region is covered with reactive metal, it is smoother and corrodes fast.
It is surrounded by acidic solution.
This corresponds to morphological experimental observations \cite{65}.
Note that for the chosen parameters, the matter loss ratio is found more than two
orders of magnitude faster in the localized regime compared to the generalized one.

\subsubsection{A common scenario}

To our knowledge, this formalism is the only one capable of depicting
generalized and localized corrosion with three distinct cases of metal
surface, pitting and occluded corrosion.

The three case studies have demonstrated that corrosion is slow at the beginning
with  a generalized corrosion regime followed by a strong acceleration of the corrosion
process of several orders of magnitude when localized corrosion settles.
Separation of half-reactions catalyzed by local different environment is
essential and material loss is substantially different in the anodic and
cathodic regions.
The cathodic passivated regions experience little material loss whereas
the anodic regions corrode fast.

The stochastic nature of the simulation is also crucial for the scenario.
It is the source of inhomogeneity on the surface: random spatial distribution of
anodic and cathodic sites.
Moreover, in the volume, the neutralization is dependent on the neutralization of the acidic and basic sites
which occurs as a random walk event in the modelling of diffusion [equation~(\ref{eq:neutralizationCA})].
This randomness can lead to spatial density inhomogeneities of acidic and basic species.
The local excesses of acidic or basic species, may sometimes be difficult
to neutralize because the neutralization takes place essentially only on the dividing surface
of a given excess of a species (for instance acidic) with the neighbouring regions of different
species (basic species for instance).
It has been shown that the kinetics of neutralization does not follow a
standard kinetic regime since it is the case when there is perfect mixing:
the inhomogeneity is slower to neutralize \cite{66,67}.
The surface and solution inhomogeneities are then coupled by the two half-reactions.

\subsection{The benefits of corrosion: in search of materials} \label{ssec:newmaterials}

We previously observed that kinetics and morphology of the surface are deeply related.

\begin{figure}[!b]
\centerline{\includegraphics[width=1.00\textwidth]{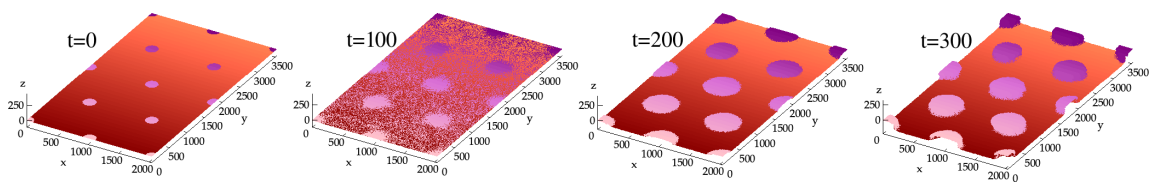}}
\caption{(Color online) Snapshots of the surface for different times as indicated in the figure from $R=100$
initial patterns. Colour code as in figure~\ref{fig:generalisedsnap}.}
\label{fig:structM1}
\end{figure}

In the bare metal case and at an occluded corrosion, the localization of the anodic
and cathodic zones is random.
In the case of pitting, the anodic region appears at the location of the
default, which is placed in the middle of the surface only for the reason of
having the pit centred in the simulation.
Here, we intend to spatially orient the corrosion by prepatterning the initial
surface into bare metal and the regions covered with oxide.
This can be achieved using stencil masks which can be realized using 
numerous lithography techniques \cite{68,69}.
These patterns have been chosen as disks of oxide of radius $R$.
As we shall consider the lattices of $2048\times 3544\times 544$ to
allow for the description of a repeating pattern, we need to use a value of  $N_{\mathrm{diff}}$ where
calculation time remains reasonable. We set $N_{\mathrm{diff}}=100$.
Such large lattices have been made accessible by using GPU computing techniques, in NVIDIA's CUDA
environment \cite{70}.
A typical initial surface is shown in figure~\ref{fig:structM1} for $t=0$.

We can observe at $t=100$, the appearance of oxide over the whole surface.
However, the initial oxide patterns are dominant and have grown in size
in comparison with their initial size.
At this stage, the patterns induce a separation between anodic and cathodic zones.
This separation becomes more pronounced at $t=200$, where one can see that the oxide
is restricted to regions around the initial patterns. The rest of the metal surface
has been cleaned, in the sense that the acidic medium in these areas hinders
the formation of oxide and supports its dissolution.
Finally, for $t=300$, due to the corrosion in the bare metal regions,
three-dimensional columns emerge from the surface.

\begin{figure}[!t]
\begin{center}
\includegraphics[width=0.5\textwidth]{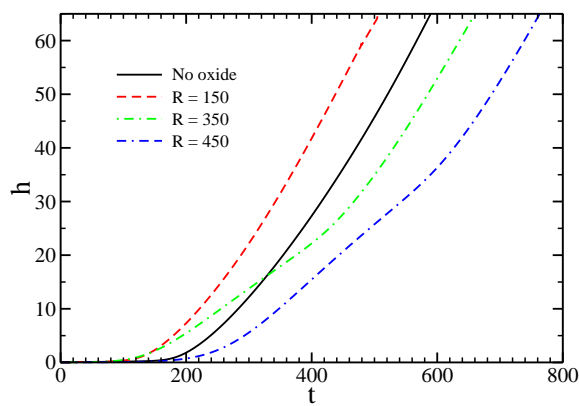}
\caption{(Color online) Average height of the surface for different initial oxide
radius.}
\label{fig:bigmaterloss}
\end{center}
\end{figure}

To study the effects of this prepatterning, we have compared different cases.
The first is again a bare metal, which serves as a reference, then
systems with disks of oxide of various radius.
The equivalent height of the layer is reported in figure~\ref{fig:bigmaterloss}.
In the order of no oxide to more oxide on the initial surface, we can observe
that the behaviour is non-monotonous.
First for $R=150$, the current with respect to the bare surface increases,
which is due to the transition to a localized corrosion which starts earlier since it
is promoted by the prepatterning.
If the size of patterns is further increased, the behaviour is reversed.
The curves are lower and can even become lower than the bare metal surface.
This can be understood due to an increasing passivation which, finally,
becomes a leading effect.

\begin{figure}[!t]
\centerline{\includegraphics[width=0.7\textwidth]{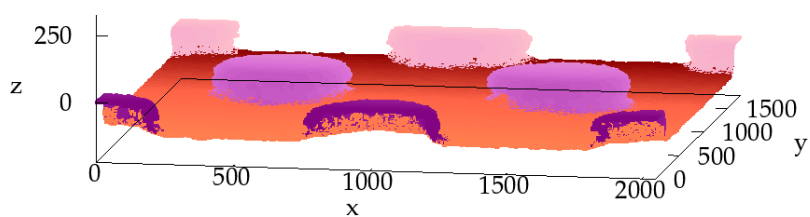}}
\caption{(Color online) Sideview of the surface for $t=300$ of the snapshots of figure~\ref{fig:structM1}. Color code as in figure~\ref{fig:generalisedsnap}.}
\label{fig:struct3D}
\end{figure}

This study has two features which are of interest.
Firstly, figure~\ref{fig:struct3D} shows that a simple surface prepatterning
may be used to orient and obtain three-dimensional structures.

Secondly, there are situations when acceleration of corrosion is needed.
This is the purpose of biodegradable materials, such as those needed in surgery
like stents or prothesis as bone reparing screws \cite{64}.
The geometrically organized surface permits a finer control with a more homogeneous corrosion.
Experiments have been conducted illustrating the galvanic effect
of depositing on the surface of noble metals such as Au or Pt \cite{69,71}
deposited on iron which exhibit a similar scenario.
Here, we propose to use the oxide itself as a natural replacement.
One point of interest is that it is produced in the course of a corrosion
process itself and will follow the morphology of the metallic interface,
whereas initially deposited metals will eventually fall off.

\section {Conclusion}

In this paper, we study a modelling of corrosion based on
spatially separated electrochemical half-reactions, diffusion,
acido-basic neutralization in solution and passive properties of the oxide layers.
The CA modelling framework is shown to be well suited to include these features.
The CA modelling extends the macroscopic deterministic approaches providing
microscopic features via the introduction of stochastic effects.
We show that by only varying the initial conditions, the same model can describe the
generalized corrosion and the localized corrosions, active and passive surfaces,
including occluded corrosion phenomena.
A common scenario emerges, where a fast localized corrosion is associated with spatial
separation of anodic and cathodic reactions.
The stochastic approach is essential to trigger the instability leading to spatial
separation of anodic and cathodic regions and, respectively, acidic and basic zones
in a solution.
Kinetic regimes and morphology appear to be closely related.
This motivated the investigation of a specific initial metal surface state
with a prepatterned oxide layer covered metal.
The study shows that the method can be used to control the material morphology
and gain a better control of the kinetics of the corrosion rate.
The model will be completed in future to include other phenomena
such as oxygen evolution or IR drop and will account for quantities like pitting factor
\cite{72,73}.
This study is highly indebted to Jean-Pierre Badiali for the
concepts and the model developed.

\section*{Acknowledgements}
M. Zenkri, F. Raouafi, F. Ben Cheikh Larbi and D. di Caprio
acknowledge the financial support by the \textit{PHC Utique} program of the French
Ministry of Foreign Affairs and Ministry of higher education and research and
the Tunisian Ministry of higher education and scientific research in the CMCU
project number 17G1207.

\ukrainianpart

\title{Вклад методу коміркового автомату в розуміння корозійних явищ}

\author{М. Зенкрі\refaddr{label1,label2,label3},
        Д. ді Капріо\refaddr{label2}, С. Перез-Брокат\refaddr{label4}, Д. Ферон\refaddr{label4,label5}, Ж. де Ламар\refaddr{label4}, А.~Шусe\refaddr{label5}, Ф. Бен Шейх Ларбі\refaddr{label1,label3}, Ф. Руафі\refaddr{label1,label3}}
\addresses{
\addr{label1} Лабораторія фізичної хімії мікроструктур і мікросистем, Інститут наукових і технічних досліджень,\\ вул. Сіді Бу Саід, B.P:51 2075 Ла Марса, Туніс
\addr{label2}  Дослідницький університет науки та літератури Парижу, Chimie ParisTech --- CNRS, Інститут хімічних досліджень Парижу, Париж, Франція
\addr{label3} Факультет наук Бізерту, 7021 Ярзуна, Університет Карфагену, Туніс
\addr{label4} Лабораторія корозії та поведінки матеріалів у середовищі, Комісаріат атомної та альтернативної енергетики (SCCME), Університет ``Париж-Сакле'', Гіф-сюр-Івет, Франція
\addr{label5} Лабораторія аналізу і моделювання біології та навколишнього середовища, дослідницький центр 8587, CNRS-CEA-Університет Еврі Валь Есон, Франція 
}

\makeukrtitle

\begin{abstract}

Представлено стохастичний метод моделювання корозії на основі розділених у просторі електрохімічних напів-реакцій, дифузії, 
лужно-кислотної нейтралізації розчину і пасивуючих властивостей оксидних шарів. Єдиний підхід дозволяє для різних початкових умов описувати узагальнену корозію, локалізовану корозію, реактивні та пасивні поверхні, в тому числі явища абсорбованої корозії. Спонтанне просторове розділення анодної та катодної зон пов'язане з чистим металом і пасивованим металом на поверхні, а також з локальною кислотністю розчину. Ця зміна є наслідком набагато більшої швидкості корозії. Морфологія матеріалу тісно пов'язана з кінетикою корозії, що може бути використано у прикладних технологіях.  
\keywords електрохімія, корозія, локалізована корозія, пасивація, комірковий автомат

\end{abstract}

\end{document}